\documentclass[conference]{IEEEtran}
\IEEEoverridecommandlockouts

\usepackage{cite} 
\usepackage{amsmath,amsfonts,amssymb}
\usepackage{algorithmic}
\usepackage{graphicx} 
\usepackage{textcomp}
\usepackage{xcolor}

\usepackage{graphics} 
\usepackage{epsfig} 
\usepackage{mathptmx} 
\usepackage{multirow} 
\usepackage{verbatim} 

\usepackage{bm}
\usepackage{color} 
\usepackage{tabularx} 
\usepackage{algorithm} 
\usepackage{diagbox}
\usepackage{float} 
\usepackage{subfigure} 

\usepackage[english]{babel}
\usepackage{times} 
\usepackage{amsbsy}
\usepackage{balance} 
\usepackage{latexsym} 
\usepackage{txfonts} 
\usepackage{wasysym}
\usepackage{mathbbol} 
\usepackage{epstopdf} 
\allowdisplaybreaks

\def\BibTeX{{\rm B\kern-.05em{\sc i\kern-.025em b}\kern-.08em
    T\kern-.1667em\lower.7ex\hbox{E}\kern-.125emX}} 

\begin{document}
\title{Resource Allocation for Non-Orthogonal Multiple Access (NOMA) Enabled LPWA Networks}
\author{Kaihan Li, Fatma Benkhelifa and Julie McCann\\
Imperial College London, London, UK\\
Email: \{k.li18, f.benkhelifa, j.mccann\}@imperial.ac.uk
}

\maketitle
\begin{abstract}
{\ In this paper, we investigate the resource allocation for uplink non-orthogonal multiple access (NOMA) enabled low-power wide-area (LPWA) networks to support the massive connectivity of users/nodes. Here, LPWA nodes communicate with a central gateway through resource blocks like channels, transmission times, bandwidths, etc. The nodes sharing the same resource blocks suffer from intra-cluster interference and possibly inter-cluster interference, which makes current LPWA networks unable to support the massive connectivity. Using the minimum transmission rate metric to highlight the interference reduction that results from the addition of NOMA, and while assuring user throughput fairness, we decompose the minimum rate maximization optimization problem into three sub-problems. First, a low-complexity sub-optimal nodes clustering scheme is proposed assigning nodes to channels based on their normalized channel gains. Then, two types of transmission time allocation algorithms are proposed that either assure fair or unfair transmission time allocation between LPWA nodes sharing the same channel. For a given channel and transmission time allocation, we further propose an optimal power allocation scheme. Simulation evaluations demonstrate approximately $100dB$ improvement of the selected metric for a single network with $4000$ active nodes.
\par}
\end{abstract}
\begin{IEEEkeywords}
	Internet-of-Things (IoT), low-power wide-area (LPWA) networks, uplink NOMA, LoRa, interference cancellation, throughput fairness, channel allocation, spreading factor (SF) allocation, power allocation.
\end{IEEEkeywords}

\section{Introduction}
{\ Internet of Things (IoT) is the core enabler for many emerging applications such as urban smart parking, industrial manufacturing, agricultural yield prediction, home automation and personal health care \cite{AlaIoT}. Many such applications span multi-kilometer distances therefore networks to support them are becoming energy-efficient covering equally long distances - an alternative to multi-hop networks due to the latter being complex to maintain and lacking robustness. This paper's focus is on these low-power wide-area (LPWA) networks \cite{UsmanLPWAN} such as SigFox, narrowband-IoT (NB-IoT), LoRa, etc. Unlike traditional cellular systems, LPWA approaches are designed to support long node lifetimes (i.e., they are inherently low-powered to provide up to $20$ years battery life for example) and they should cover longer ranges up to tens of kilometers. The majority of applications in this field are delay-tolerant which facilitates the aforementioned distance benefit which is achieved through lower data rates down to hundreds of bits per second. LoRa, SigFox, Weightless, Ingenu, etc. \cite{UsmanLPWAN,ZhijinResourceAllocation} mostly occupy the sub-1GHz unlicensed industrial, scientific, and medical (ISM) spectrum like $868$ MHz in Europe. Their counterparts are the cellular based solutions that operate over the licensed spectrum such as narrow-band IoT (NB-IoT) \cite{3GPPNBIoT} and enhanced machine-type communication (eMTC) which benefit from legacy GSM and LTE \cite{LimitsofLoRaWAN}. This paper focus on the former class of LPWA network.
\par}
{\ LPWA technologies face challenges with respect to their coexistence and scalability. \cite{SubGHzLPWAN} points out three co-existence and interference issues grouped as: cross-technology issues for different technologies in the same unlicensed ISM band, intra-technology/inter-network issues for different networks using one same technology, and intra-network issues for different nodes within one single network.
Current LPWA solutions are susceptible to interference and do not even handle intra-networks well. For example, LoRa (LoRaWAN) evaluations show that a collision avalanche occurs with a high number of nodes and the systems' capacity degrades quickly when the load on the link increases \cite{OntheLimitsofLoRaWANChannelAccess}. \cite{CanLoRaScale} investigates the co-spreading factor interference (co-SF) for LoRa that occurs when concurrent signals reside in the same SF and frequency, which results in an exponential drop of coverage probability as the number of nodes increases. Preliminary deployments and experiments in \cite{DoLoRaScale} shows that only $120$ nodes can be supported per $3.8$ hectare for a typical deployment. These practical results are far from expectations indicating that the current LPWANs cannot scale to support massive numbers of nodes due to interference. 
\par}
{\ Non-orthogonal multiple access (NOMA) permits signal overlaps in either of time or frequency exploiting the power domain or code domain to increase the numbers of multiple users/nodes sharing the same resource block (e.g., time, channel/frequency, bandwidth, etc.) for cellular networks. As a power-domain user multiplexing technology, NOMA adopting a superposition coding (SC) transmission and successive interference cancellation (SIC) is able to decode the signals sequentially through the exploitation of different power levels \cite{YuyaNOMA}. Compared to the traditional orthogonal multiple access (OMA) techniques, NOMA's potential to support massive connectivity and to enhance spectral efficiency has been shown in Heterogeneous networks (HetNets) together with massive multiple-input multiple-output (MIMO) \cite{YuanweiNOMA}. A power control method for uplink NOMA is proposed in \cite{NingboUplinkNOMA} that surpasses the OMA solution in terms of the sum rates. NOMA has also been explored for more general uplink data transmissions as well as downlink in \cite{HinaUserClusteringNOMA} where user clustering and power allocation are jointly designed for sum throughput maximization. \cite{FatmaLoRaEnergyHarvesting} focuses on the uplink data transmissions of LoRa network. \cite{MichaNOMAcellularIoT} highlights the possibility of NOMA for cellular based IoT.
\par}
{ In this paper, we propose to maximize the minimum uplink transmission rate of LPWA nodes where NOMA is adopted at a single gateway. The major contributions of this paper can be summarized as:
\begin{enumerate}
  \item As far as we are aware, we are the first to apply and model uplink power-domain NOMA to LPWA networks which allows multiple LPWA nodes sharing the same resource blocks to communicate under much less interference or even interference-free therefore supporting massive connectivity.  
  \item We formulate the resource allocation of uplink NOMA-enabled LPWA network as a minimum transmission rate maximization problem to improve connectivity while guaranteeing user fairness considering constraints in real implementations. We understand that this type of formulation is new to cellular and LPWA networks.
  \item We decompose the optimization problem into three sub-problems: nodes clustering channel allocation, transmission time allocation (equivalent to SF allocation for LoRa networks), and power allocation. The corresponding proposed algorithms make the optimization problem tractable and have demonstrated a good trade-off between system performance and computational complexity.
  \item In simulation, LoRa network is considered as a practical example of LPWA networks. The proposed NOMA enabled LPWAN and resource allocation schemes together demonstrates significant performance improvement.
\end{enumerate}
\par}
\section{System Model}\label{S2}
{\ In this paper, we consider the uplink transmissions in a LPWA network. We assume a single gateway located at the center of a circle with radius $r$ surrounded by randomly distributed $N$ nodes. The distance from the node $U_n$ to the central gateway is denoted by $d_n$ for $n={1,...,N}$. Both the gateway and the node are equipped with a single antenna. Every LPWA node communicates with the central gateway through different resource blocks (e.g., time, channel/frequency, bandwidth, etc.) determined by the LPWA network. Here we concentrate on the resource blocks of time and channel, which are applicable to many LPWA networks, to maintain generality. For example, despite of multiple channels, LoRa also have different spreading factors which lead to different time-on-air/transmission times.
Similarly, NB-IoT transmit repeated preambles over multiple sub-carriers for different number of repetitions \cite{NanJiangRACHNBIoT}. 
\par}

\subsection{Nodes sharing the same resource blocks}
{\ The nodes sharing the same resource blocks form a cluster. For now, we assume that we have intra-cluster and inter-cluster interference since some LPWA networks do not assure orthogonality between resource blocks. For instance, LoRa networks have been criticized for their imperfect orthogonality \cite{CroceInterSF} between SFs which represent different transmission times.
\par} 
{\ We denote $\Omega_w$ the cluster of LPWA nodes sharing the same resource blocks as nodes transmitting over the same channel $Q_w$ with the same bandwidth $B_w$ during the same transmission time $T_w$. First, we assume that all nodes start transmitting at the same time. Also, the number of nodes per cluster is unpredictable while it is upper bounded by the total number of nodes $N$. The total number of clusters $\Omega_w$ is equal to $W$. In addition, the total number of channels and transmission times are equal to $K$ and $F$ respectively and we will have $W=K\times F$. Furthermore, we let $\mathbb{\Omega}=\{\Omega_w, w=1,\dots, W\}$ being the set of all possible clusters with size $W$ grouping all LPWA nodes. Then, we have:
\begin{itemize}
    \item The clusters sharing the same channel (i.e., $Q_w=Q_k$) are grouped under the subset $\mathbb{Q}_k=\bigcup\{\Omega_w \text{ s.t. } Q_w=Q_k \}$ for each $k$ of $k=1,\dots,K$. Each subset $\mathbb{Q}_k$ is disjoint with the others. All of these subsets together form the union $\mathbb{\Omega}_K$ where $\mathbb{\Omega}_K=\bigcup_{k=1}^K \mathbb{Q}_k$ and $\mathbb{Q}_k \bigcap\mathbb{Q}_{k'}= \emptyset$ for $k\neq k'$. 
    \item Similarly, the clusters sharing the same transmission time $T_f$ are grouped under the subset $\mathbb{S}_f=\bigcup \{\Omega_w \text{ s.t. } T_w=T_f \}$ for each $f$ of $f=1,\dots,F$. Each subset $\mathbb{S}_f$ is disjoint with the others. All of these subsets together form the union $\mathbb{\Omega}_F$ where $\mathbb{\Omega}_F=\bigcup_{f=1}^F \mathbb{S}_f$ and $\mathbb{Q}_f \bigcap\mathbb{Q}_{f'}= \emptyset$ for $f\neq f'$. 
\end{itemize}
Each node can be allocated at most to one cluster at the same time. Each cluster belongs to only one subset $\mathbb{Q}_k$ and only one subset $\mathbb{S}_f$ (i.e., $\exists! k=1,\dots,K,\ \exists! f=1,\dots,F$, s.t. $\Omega_w \in \mathbb{Q}_k \cap \mathbb{S}_f$). We use the binary variable $\xi_{w,n}\in\{0,1\}$ to indicate whether the node $U_n$ belongs to the cluster $\Omega_w$:
\begin{align}
    \xi_{w,n} &=\begin{cases}
    1, &\mbox{ if } n \in \Omega_w,\\
    0, &\mbox{otherwise}.
    \end{cases}
\end{align}
Moreover, the binary variable $\xi^S_{f,n}\in\{0,1\}$ is used to indicate whether the node $U_n$ transmits on the time $T_f$ where $f=1,\dots,F$. The binary variable $\xi^Q_{k,n}\in\{0,1\}$ indicates whether the node $U_n$ is transmitting over the channel $Q_k$ where $k=1,\dots,K$. Hence, the result of their multiplication (i.e., $\xi_{w,n}=\xi^S_{f,n}\times\xi^Q_{k,n}$) indicates whether the node $n$ belongs to the cluster $\Omega_w$ occupying the channel $Q_k$ and having the transmission time $T_f$. Obviously, any LPWA node $i$ with the same transmission time $T_f$ as the node $n$ verifies $\xi^S_{f,n}=\xi^S_{f,i}=1$ while $\xi^S_{f,n}=1- \xi^S_{f,i}$ for those nodes having different transmission times from $T_f$. 
\par}
\subsection{Intra-Cluster and Inter-Cluster Interference Scenarios}
{\ The interference among LPWA nodes at the gateway depends on the resource blocks they are occupying. The nodes belonging to the same cluster interfere with each other where $\xi_{w,n}=\xi_{w,i}$ for $i\neq n$, namely intra-cluster interference. The nodes belonging to different clusters may interfere with each other where $\xi_{w,n}\neq \xi_{w,i}$ for $i\neq n$. No interference occurs if their clusters belong to different subsets $\mathbb{Q}_k$ (i.e., $\xi^Q_{k,n}\neq \xi^Q_{k,i}$ for $i\neq n$). We assume that nodes occupying different channels do not interfere with each other. However, if their clusters belong to the same subset $\mathbb{Q}_k$ while the different subset $\mathbb{S}_f$, they interfere with each other which we name it as inter-cluster interference and the impact of interference depends on which transmission time is used. The inter-cluster interference happens only when the nodes share the same channel and different transmission times (i.e., $\xi^Q_{k,n}=\xi^Q_{k,i}$ and $\xi^S_{f,n}\neq \xi^S_{f,i}$ for any $i\neq n$). Take LoRa as an example, the use of different SFs which represents distinct transmission times does not guarantee interference-free communication (i.e., imperfect orthogonality of SFs). 
\par}
\subsection{Physical Channel Modeling}
{\ As for the channel model, we consider the urban and suburban areas where most likely there are no dominant line-of-sight (LoS) component between the LPWA nodes and gateway. Therefore, the channel between an arbitrary node $U_n$ and the central gateway is modeled as a Rayleigh fading channel with path loss. The extensively used Log-distance propagation model has been chosen as our path loss model for LPWAN. Accordingly, the channel gain $g_{w,n}$ for the node $U_n$ assigned to the cluster $\Omega_w$ can be expressed as follows:
    \begin{align}
       g_{w,n}=\eta_{w,n} h_{w,n} d_n^{-\beta},     \label{channelgain}
    \end{align}
where $\eta_{w,n}$ is a constant related to the path loss, $h_{w,n}\thicksim exp(1)$ represents the small-scale fading which is exponentially distributed with the unit mean, and $\beta\in\left[3,5\right]$ is the Log-distance path loss exponent for shadowed urban area. We assume the channel state information (CSI) is known at the gateway.
\par}
{\ Let us consider a desired node $U_n$ belonging to cluster $\Omega_w$ transmitting during time $T_f$ over channel $Q_k$ (i.e., $\Omega_w \in \mathbb{S}_f\cap\mathbb{Q}_k$). Without interference cancellation at the central gateway, the signal-to-interference-plus-noise ratio (SINR) $\varphi_{w,n}$ of node $U_n$ is given by:
\begin{align}
    \varphi_{w,n}=\frac{ p_{w,n} g_{w,n}}{\mathcal{I}_{intra\!\_\!w,n}+ \mathcal{I}_{inter\!\_\!w,n}+\sigma_w^2}, \label{SINR}
\end{align} 
where $\mathcal{I}_{intra\!\_\!w,n}=\sum_{i \in \Omega_w, i\ne n} \xi_{w,i} p_{w,i} g_{w,i}$ is the intra-cluster interference. $\mathcal{I}_{inter\!\_\!w,n}=\sum_{i\in \Omega_{w'}, w'\neq w} \xi_{w',i} col_{i,n} p_{w',i} g_{w',i}$ is the inter-cluster interference only for nodes sharing the same channel and transmitting during different times where $\xi_{w',i}=\left(1-\xi^S_{f,i}\right) \xi^Q_{k,i}$.
$p_{w,n}$ is the uplink transmission power for node $U_n$ over the channel $Q_k$ and $i$ represents the LPWA interfering node $U_i$. $\sigma_w^2$ is the variance of the zero-mean additive white Gaussian noise (AWGN) over the channel $Q_k$ with bandwidth $B_w$. In this paper, $col_{i,n}=\frac{\min(T_f,T_i)}{T_f}$ is the collision time between nodes $U_n$ and $U_i$ with different transmission times $T_f$ and $T_i$ respectively. The intra-cluster and inter-cluster interference are the main reasons for the performance degradation.
\par}


\section{Max-Min Rate Optimization}\label{S3}
{\ As illustrated, the interference between LPWA nodes will occur within the same cluster sharing the same resource blocks (namely intra-cluster interference), and between the clusters sharing the same channel (namely inter-cluster interference). NOMA utilizes an additional domain (i.e., power domain) to superpose multiple nodes sharing the same resource blocks (e.g., time, channel, frequency, code, spreading factor, etc.). The separation of these nodes at the receiver is achieved through successive interference cancellation (SIC) and capacity-achieving channel codes such as the Turbo code and low-density parity check (LDPC) code which allow the massive connectivity for LPWA networks. 
Robust multiple access can be achieved by executing SIC at the gateway for LPWA nodes sharing the same resource blocks (e.g., channel, transmission time).
\par}
{\ Considering the NOMA-enabled LPWA for uplink transmissions, the desired node $U_n$ inside the cluster $\Omega_w$ transmitting over the channel $Q_k$ with bandwidth $B_w$ and transmission time $T_f$ will be interfered by the nodes from the same cluster $\Omega_w$ and from the clusters $\Omega_w'$ ($w'\neq w$) having the same channel $Q_k$ and different transmission times $T_f'$ ($f\neq f'$). Before applying the SIC, the LPWA nodes interfering with the others (i.e., having the same bandwidth and any transmission time) are ordered according to the decreasing channel gains normalized by the noise power (i.e., $\gamma_{w,n}=g_{w,n}/\sigma_w^2$ for $w=1,\dots,W$). For ease of notation, we assume that $\gamma_{w,1}>\gamma_{w,2}>\gamma_{w,3}>\dots$ for the cluster $\Omega_w$ where $w=1,\dots,W$. Firstly, the gateway will decode the signal corresponding to the most powerful channel gain and then subtract its component from the received signal before decoding the signal corresponding to the second  powerful channel gain. The gateway successively subtracts the signal corresponding to the more  powerful channel gain before decoding the signal corresponding to the following powerful channel gain, until the last node in the cluster is decoded. 
The reason for this ordering is that the initial decoded signal has to endure the interference from the others while the latter decoded signal can benefit the throughput by canceling the interference which should have been taken by the previously decoded stronger signals. 
In other words, nodes or end-devices with better channel gains will be decoded first to achieve satisfactory rate even under the existence of stronger interference while the latter decoded nodes with worse channel gains will normally experience less interference.
\par}
{\ Applying the NOMA-enabled LPWA for uplink transmissions, the SINR $\varphi_{w,n\_NOMA}$ for node $U_n$ in $\Omega_w$ occupying the channel $Q_k$ is denoted as:
\begin{align}
    \varphi_{w,n}^{NOMA}=\frac{ p_{w,n} g_{w,n}}{\mathcal{I}^{NOMA}_{intra\!\_\!w,n}+ \mathcal{I}^{NOMA}_{inter\!\_\!w,n}+\sigma_w^2}, \label{SINR_NOMA}
\end{align} 
where $\mathcal{I}^{NOMA}_{intra\!\_\!w,n}=\sum_{i \in \Omega_w, i\ne n} \xi_{w,i}\mu_{w,i}  p_{w,i} g_{w,i}$ and $\mathcal{I}^{NOMA}_{inter\!\_\!w,n}=\sum_{i\in \Omega_{w'}, w'\neq w} \xi_{w',i} \mu_{w',i}  col_{i,n} p_{w',i} g_{w',i}$ are the intra-cluster interference and the inter-cluster interference respectively. The latter is only for nodes sharing the same channel $Q_k$ while transmitting during different times (i.e., $\xi_{w',i}$ verifies $\left(1-\xi^S_{f,i}\right) \xi^Q_{k,i}$). $\mu_{w,i}=1$ indicates $\gamma_{w,n}>\gamma_{w,i}$ and $0$ otherwise. This binary variable $\mu_{w,i}$ will not bring new constraints by the virtue of the SIC of NOMA. The detail of the implementation of SIC can be found in \cite{HinaUserClusteringNOMA}. 
The transmission rate of $n^{th}$ node over the $w^{th}$ cluster after the use of SIC by the virtue of NOMA is given by:
     \begin{align}
        R_{n}^{NOMA} &= \sum_{w=1}^W B_w \xi_{w,n} \log_2(1+\varphi_{w,n}^{NOMA}). 
    \end{align} 
\par}
{\ To improve the connectivity and to guarantee user fairness, we propose to maximize the minimum transmission rate of the nodes after using SIC at the gateway while optimizing channel, transmission time and power allocation. Without any loss of generality, we formulate the problem with a close focus on LoRa networks where transmission times are directly related to the assignment of SFs. The minimal transmission rate maximization problem is formulated as follows: 
\begin{subequations}\label{formulated}
\begin{align}
\left( \textbf{P1} \right)\;\;
&\mathop {\max }\limits_{{\xi_{w,n}}} \; {{\mathop {\min } \sum_{w=1}^W B_w \xi_{w,n} \log_2(1+\varphi_{w,n}^{NOMA}) }}\\
\text{subject to}\;\;
&C_1:\; P_{min} \leq p_{w,n} \leq P_{max},\;\forall \;w,\;n,\label{3a}\\
&C_2:\;p_{w,n} g_{w,n}\geq \theta_{f},\;\forall \;w,\;n,\;f, \label{3b}\\
&C_3:\;p_{w,n}g_{w,n}\geq p_{w,n-1} g_{w,n-1},\forall w,n\ne 1, \label{10b}\\
&C_4:\;\xi_{f,n}^{S }\in\{0,1\}, \;\forall \;f,\;n, \label{2aa}\\
&C_5:\;\xi_{k,n}^{Q}\in\{0,1\}, \;\forall \;k,\;n, \label{2bb}\\
&C_6:\;\xi_{w,n}=\xi_{f,n}^{S} \xi_{k,n}^{Q}\in\{0,1\}, \;\forall \;w,\;n, \label{1a}\\
&C_7:\;\sum_n \xi_{w,n}\leq 1,\;\forall \;w ,\label{1b}
\end{align}
\end{subequations}
where the constraint $C_1$ limits the transmission power for all LPWA nodes being within the maximum and minimum values, the constraint $C_2$ assures that the signal can be successfully decoded on the premise of satisfying receiver's sensitivity threshold $\theta_f$. The subscript $f$ here implies the threshold is also related to the spreading factors in LoRa except for the transmission time. The constraint $C_3$ follows the order of uplink NOMA. The constraint $C_4$ shows that $\xi^S_{f,n}$ can be either $0$ or $1$ for each node and transmission time $T_f$ pair where $f=1,\dots,F$. The constraint $C_5$ shows that $\xi^Q_{k,n}$ can be either $0$ or $1$ for each node and channel $Q_k$ pair where $k=1,\dots,K$. The constraint $C_6$ shows that $\xi_{w,n}$ can be either $0$ or $1$ for each node and cluster pair since one node is either belonging to the specific cluster or not at the same time. The constraint $C_7$ means that the node $U_n$ cannot be allocated to more than one cluster simultaneously. 
\par}

{\ The formulated problem (\textbf{{P1}}) is a non-convex mixed-integer nonlinear programming (MINLP) problem which is NP-hard and exhaustive search yields exponential time complexity. Thus, we propose to decouple (\textbf{{P1}}) into three sub-problems where we optimize the channel, transmission time and power allocation separately and gradually. We firstly assign the channels to the LPWA nodes following by the transmission times. These two sub-problems are performed assuming that all LPWA nodes are transmitting with their maximum transmit power. We then optimize the transmit power for the LPWA nodes. The proposed resource allocation schemes make the objective function (i.e., the max-min problem of the NOMA enhanced LPWA network) tractable by releasing the binary and correlated resource blocks constraints. Furthermore, the desired network performance metric has been significantly improved after the optimization.
\par}
\begin{algorithm}[ht]
\caption{Algorithm Summary for Solving (\textbf{P1})}
\label{csp_algorithm}
\begin{algorithmic}[1]
\renewcommand{\algorithmicrequire}{\textbf{Input:}} 
\REQUIRE $W$,\ \!$N$,\ \!$K$,\ \!$d_n$,\ \!$g_{k,n}$,\ \!$\sigma_k^2$,\ \!$\gamma_{k,n}$,\ \!$F$,\ \!$T_f$,\ \!$r$,\ \!$B_w$,\ \!$P\!_{max}$,\ \!$P\!_{min}$,\ \!$\epsilon$,\ \!$\theta_f$;\\
\STATE Sort $\gamma_{w,n}$ in $\tilde{\gamma}_m= \text{sort}(\gamma_{w,n}, \text{`descending'})$, $m=1,\dots, N$;
\IF {$\!N\!\!\!\!\!\mod\! K=0$}
\STATE $1^{st}\ channel = \{\tilde{\gamma}_1,\tilde{\gamma}_{K+1},\tilde{\gamma}_{2K+1},\dots,\tilde{\gamma}_{N-K+1}\}$;\\
\STATE $2^{nd}\,channel = \{\tilde{\gamma}_2,\tilde{\gamma}_{K+2},\tilde{\gamma}_{2K+2},\dots,\tilde{\gamma}_{N-K+2}\}$;\dots;\\
\STATE $K^{th}\,channel = \{\tilde{\gamma}_K,\tilde{\gamma}_{2K},\tilde{\gamma}_{3K},\dots,\tilde{\gamma}_{N}\}$; find $N_k$ for all $k$;
\ELSE
\STATE $q=\!N\!\!\!\!\!\mod\! K$, do lines $3-5$ for the first $N\!-\!q$ nodes, distribute rest $q$ nodes to first $q$ channels, update $N_k$;
\ENDIF
    \FOR{k=1:K}
    \STATE {\bf if} Unfair {\bf then} $N_k^f=\frac{N_k}{F}$ ($f$=1:F), $z=N_k\!\!\!\!\!\mod\! F$;\\ 
    \STATE \ \ \ {\bf if} $z\ne 0$, distribute rest $z$ nodes to first $z$ groups;\\
    \STATE \ \ \ Assign $T_{f'}$ ($f'$=F:1) to $N_k^f$ nodes in $N_k$ head to tail;\\
    \STATE {\bf if} Random {\bf then} each node randomly chooses a $T_f$;\\
    \STATE {\bf if} Fair {\bf then} $N_k^f= \frac{N_k}{T_f }/\sum_{i=1}^F \frac{1}{T_i}$, $j=N_k\!-\!sum(round(N_k^f))$;\\
    \STATE \ \ \ {\bf if} $j\!<\!0$, $N_k^f\!=\!round(N_k^f)\!-\!1$ for those $N_k^f\!-\!\lfloor\!N_k^f\!\rfloor\!>\!0.5$;\\
    \STATE \ \ \ {\bf if} $j\!\geq\!0$, $N_k^f=\!round(N_k^f)\!+\!1$ for $j$ largest $N_k^f\!\!-\!\!\lfloor\!N_k^f\!\rfloor$;\\
    \STATE \ \ \ $N_k^f=round(N_k^f)$ for the else;\\
    \STATE \ \ \ Assign $T_f$ to $N_k^f$ nodes in $N_k$ from head to tail;\\
    \STATE {\bf if} Distance {\bf then} each chooses a $T_f$ that $\frac{(f\!-\!1)r}{F}<d_n\leq \frac{fr}{F}$;\\
    \ENDFOR
\FOR{k=1:K}
\FOR{$n=\{k,K\!+\!k,2K\!+\!k,\dots,N\!-\!K\!+\!k$\}}
\STATE $\tau_l=0$, $\tau_u=B_w log_2(1+\frac{p_{max}max(g_{k,n})}{\sigma_w^2})$;\\
\STATE Calculate $\mathcal{I}^{NOMA}_{intra\!\_\!k,n}$ and $\mathcal{I}^{NOMA}_{inter\!\_\!k,n}$;\\
\STATE {\bf while} $\tau_u-\tau_l\geq \epsilon$ {\bf do} $\varphi=(\tau_u+\tau_l)/2$;\\
\STATE \ \ \ Find $p_{k,n}$ subject to $C_1$-$C_3$ and (\ref{10d}) by cvx;
\STATE \ \ \ {\bf if} Feasible {\bf then} $p_{k,n}^{opt}=p_{k,n}$, $\tau_l=\tau$;
\STATE \ \ \ \ \ $R_{k,n}^{min}\!=\!min(B_w \log_2(1\!+\!\frac{p_{k,n}^{opt} g_{k,n}}{\mathcal{I}^{NOMA}_{intra\!\_\!k,n}+ \mathcal{I}^{NOMA}_{inter\!\_\!k,n}+\sigma_w^2 }) \forall n\!\!\in\!\! k)$;
\STATE \ \ \ {\bf if} Unfeasible {\bf then} $\tau_u=\tau$;
\STATE {\bf end} {\bf while}
\ENDFOR
\ENDFOR
\end{algorithmic}
\end{algorithm}

\subsection{Channel Allocation}
{\ A low-complexity sub-optimal channel allocation method is proposed for NOMA-LPWA network maximizing the gap between different channel gains normalized by the noise power in every channel. The signal with larger $\gamma_{w,n}$ is decoded first and its interference impact is subtracted from the following signals, while the latter decoded nodes with worse channel gains experience less interference. The purpose of doing so is to achieve satisfactory rate for stronger signals even with stronger interference while the weaker signals can benefit from the weak interference. The proposed channel allocation is formulated to distinguish the channel gains of the nodes sharing the same channel to further stress the benefit from the SIC implementation at the gateway. 
$N$ LPWA nodes will be sorted based on their normalized channel gain $\gamma_{w,n}$ in the descending order regardless of their initial occupied channels. The first $K$ nodes with the most powerful normalized channel gains are distributed among the total $K$ channels. The second $K$ nodes with the next most powerful channel gains are also distributed among the total $K$ channels, and so on. The $k^{th}$ channel among the total $K$ channels will pick the LPWA nodes whose normalized channel gains falling into the ranks $k$, $k+K$, $k+2K$, \dots. The number of nodes for each channel among $K$ available channels is decided by $N/K$. The number of nodes is exactly $N/K$ if $N\!\!\!\!\mod K = 0$. If $N\!\!\!\!\mod K\ne0$, the first $N\!\!\!\!\mod\! K$ channels will have one extra node than the last $K-N\!\!\!\!\mod\! K$ channels. Let us denote by $N_k$ the number of LPWA nodes of the $k^{th}$ channel for all $k=1,\dots,K$, which satisfies $N=\sum_{k=1}^K N_k$. If $N\!\!\!\!\mod K = 0$, we have $N_k=N/K$. Otherwise, by denoting $N\!\!\!\!\mod K$ as $q$, we have $N_k=\lfloor\frac{N}{K}\rfloor+1$ for $k=1,\dots,q$ and $N_k=\lfloor\frac{N}{K}\rfloor$ for the rest channels where $\lfloor\ \rfloor$ represents the floor function.
\par}

\begin{figure}[t]
\centering
\includegraphics[scale=0.32]{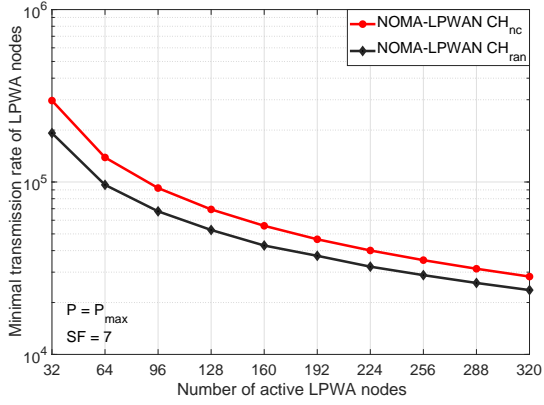}
\caption{Minimum transmission rate of nodes versus number of active nodes for NOMA-LPWAN with different channel allocation schemes.}
\label{fig1}
\end{figure}
\subsection{Transmission Time Allocation}
{\ Next, we investigate the transmission time allocation for each channel $k=1,\dots,K$ and we propose two types of allocation methods that either assign fairly or unfairly the $F$ transmission times to the LPWA nodes sharing the same channel. The unfair transmission time allocation algorithm equally allocates $F$ transmission times to the $N_K$ LPWA nodes in the channel. The fair transmission time allocation aims to ensure the interference fairness of the inter-cluster interference. Note that a larger $T_f$ makes the symbol to be more likely to collide with the others and generally less resilient to the interference. Thus, the fair transmission time allocation aims to satisfy the equality $N_k^f\times T_f=N_k^i\times T_i$ valid $\forall f,i=1,\dots,F$ where $N_k^f$ refers to the number of nodes being allocated with the transmission time $T_f$ in the $k^{th}$ channel. Therefore, the number of nodes for each transmission time $T_f$ can be derived as $N_k^f= \frac{N_k}{T_f }/\sum_{i=1}^F \frac{1}{T_i}$ which gives the smaller $N_k^f$ for the larger $T_f$.

\subsection{Power Allocation} 
{\ The global solutions cannot be obtained as (\textbf{P1}) is still nonconvex. Thus, we further propose an optimal power allocation scheme considering the power related constraints of the objective function (\textbf{P1}). By introducing a new optimization variable $\tau$, we transform the optimization problem to an equivalent convex problem finding the optimal power $p_{w,n}$ subject to constraints $C_1$-$C_3$ and a new constraint $R_{w,n}^{NOMA}\geq \tau$: 
\begin{equation}
p_{w,n} g_{w,n}\geq (2^{\frac{\tau}{B_w}}-1)(\mathcal{I}^{NOMA}_{intra\!\_\!w,n}+ \mathcal{I}^{NOMA}_{inter\!\_\!w,n}+\sigma_w^2),\;\forall\;w,\;n. \label{10d}
\end{equation}
where we amplify both sides by $10^{11}$ to get an accurate optimization result. We fix $\tau$ such that $\tau_l\leq \tau\leq \tau_u$ and then solve $p_{w,n}$ for a given $\tau$ with accuracy $\epsilon=10^{-6}$. The optimal $\tau$ can be obtained using the one-dimensional search method. Key steps are summarized in Algorithm \ref{csp_algorithm}.

\begin{figure}[t]
\centering
\includegraphics[scale=0.32]{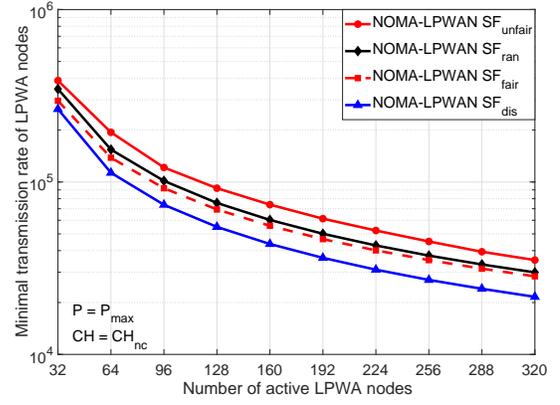}
\caption{Minimum transmission rate of nodes versus number of active nodes for NOMA-LPWAN with different transmission time/SF allocation schemes.}
\label{fig2}
\end{figure}

\section{Numerical Results}\label{S4}

{\ We simulate LoRa as an example of the LPWA network. We assume that LPWA nodes are randomly distributed around the gateway within a coverage radius $r=1km$. The number of channels $K$ is set to be $8$ and each channel operates at $868MHz$. We fix the bandwidth $B_w$ to be $125KHz$ for all clusters. The path loss exponent $\beta$ is $3.5$ and AWGN noise variance is denoted by $\sigma_w^2=-174+10\log_{10}B_w+N\!\!F$ under the room temperature where $N\!\!F=6dB$ is the typical noise figure for LoRa receiver. LoRa Chirp Spread Spectrum (CSS) modulation introduces six spreading factors (SFs) $\alpha_f=7,\dots,12$ (i.e., $f=1,\dots,6$ and $F=6$). Each LoRa symbol/chirp contains $\alpha_f$ number of bits with symbol period $\frac{2^{\alpha_f}}{B_w}$. Thus, the transmission time given by $T_f=
\frac{b}{\alpha_f}\times\frac{2^{\alpha_f}}{B_w}$ is related to the spreading factor where $b$ is the number of bits transmitted same for all $f$ for fairness. The imperfect orthogonality between SFs is taken into consideration as the result of transmission time collisions. The minimum and maximum transmit power are $0dBm$ and $20dBm$ respectively. The LoRa receiver sensitivity threshold $\theta_{f}$ for $\alpha_f$ is calculated by $\theta_{f} = \sigma_w^2+S\!\!N\!\!R_{f}^{de}$ where $S\!\!N\!\!R_{f}^{de}$ represents the required signal-to-noise ratio (SNR) for demodulation of $\alpha_f$ at LoRa SX1272/73 receiver. 
\par}

{\ Fig. \ref{fig1} illustrates how the minimum transmission rate varies with the number of active LPWA nodes for different channel allocation methods. Here we fix the transmission power to be the maximum and the SF to be $7$ for fair comparison. It can be observed that the proposed low-complexity sub-optimal nodes clustering $C\!H_{nc}$ can achieve more than 50\% improvement compared to the random channel allocation method for NOMA enabled LPWAN, since the differences between channel gains in each channel has been maximized to benefit NOMA afterwards.

Fig. \ref{fig2} shows the minimum transmission rate versus the number of active LPWA nodes for different transmission time/spreading factor allocation methods. We fix the transmission power and channel allocation method in this simulation. The unfair method gives the best performance of NOMA-LPWAN
as the result of making sure the collision times of each node's interference nodes are either smaller or equal to its own transmission time, while the fair method only guarantees the collision time fairness for all nodes occupying the same channel. The random allocation method 
may have these two characteristics only for some nodes but does not guarantee anything. The distance based $S\!\!F_{dis}$ does not follow the sequence of nodes given by $C\!H_{nc}$ when allocating.

\begin{figure}[t]
\centering
\includegraphics[scale=0.32]{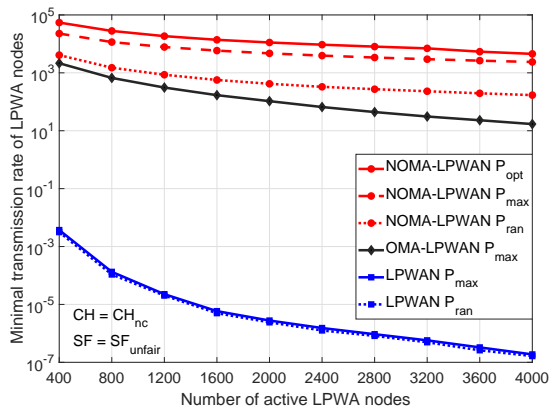}
\caption{Minimum transmission rate of nodes versus number of active nodes for NOMA-LPWAN, OMA-LPWAN and the baseline with different power allocation.}
\label{fig3}
\end{figure}

{\ Fig. \ref{fig3} shows NOMA-LPWAN's improvements, while varying numbers of nodes sharing the same resource blocks as well as power allocation methods but fixing $C\!H_{nc}$ and $S\!\!F_{unfair}$. The descending rates reveals the increasing interference. Unlike NOMA where nodes sharing the same resource blocks are superposed in power domain, we also proposed OMA-LPWAN for comparison that each node can only transmit during its own time slot within $N$ slots, which has worse performance and changes current LoRaWAN essentially. The proposed optimal power allocation with higher interference resilience outperforms the others and extends the battery life.
\par}
\section{Conclusion}\label{S5}
{\ We demonstrate that uplink non-orthogonal multiple access (NOMA) can significantly enhance low-power wide-area (LPWA) networks by alleviating interference levels of nodes sharing the same resource blocks. We decompose the minimum transmission rate maximization problem into three sub-problems consisting of channel, transmission time and power allocation. A low-complexity sub-optimal nodes clustering method is proposed followed by the unfair/fair transmission time allocation schemes and an optimal power allocation algorithm. Simulation results demonstrate that the proposed NOMA enabled network has significant potential to support massive connectivity of IoT. The other metrics will be analyzed as well as the other LPWAN examples. The test-bed based on LoRa Semtech, NOMA DOCOMO and MUIC MediaTek chipset would be able to validate NOMA-LoRa in real-world implementations.
\par}

\bibliography{LoRa_NOMA}
\bibliographystyle{IEEEtran}

\end{document}